# Metabolomics in the Cloud: Scaling Computational Tools to Big Data

Jianliang Gao[1]*, Noureddin Sadawi[1]*, Ibrahim Karaman[2], Jake T M Pearce[1], Pablo Moreno[3], Anders Larsson[4], Marco Capuccini[4], Paul Elliott[2], Jeremy K Nicholson[1], Timothy M D Ebbels[1]** and Robert Glen[1,5]**

1. Division of Integrative Systems Medicine and Digestive Disease, Department of Surgery and Cancer, Imperial College London, Sir Alexander Fleming Building, South Kensington, LondonSW7 2AZ, United Kingdom
2. Department of Epidemiology and Biostatistics, Medical Research Council–Public Health England (MRC-PHE) Centre for Environment and Health, School of Public Health, Imperial College London, London W2 1PG, UK.
3. European Molecular Biology Laboratory, European Bioinformatics Institute (EMBL-EBI), Wellcome Genome Campus, Hinxton, Cambridge CB10 1SD, United Kingdom.
4. Systematic Biology, Department of Organismal Biology, Evolutionary Biology Centre, Uppsala University, Uppsala 75236, Sweden
5. Department of Chemistry, Centre for Molecular Informatics, University of Cambridge, Lensfield Road, Cambridge, CB21EW, UK

*Contributed equally.
**Corresponding authors: t.ebbels@imperial.ac.uk, r.glen@imperial.ac.uk,

# Abstract

**Background** Metabolomics datasets are becoming increasingly large and complex, with multiple types of algorithms and workflows needed to process and analyse the data. A cloud infrastructure with portable software tools can provide much needed resources enabling faster processing of much larger datasets than would be possible at any individual lab. The PhenoMeNal project has developed such an infrastructure, allowing users to run analyses on local or commercial cloud platforms. We have examined the computational scaling behaviour of the PhenoMeNal platform using four different implementations across 1-1000 virtual CPUs using two common metabolomics tools.

**Results**: Our results show that data which takes up to 4 days to process on a standard desktop computer can be processed in just 10 min on the largest cluster. Improved runtimes come at the cost of decreased efficiency, with all platforms falling below 80% efficiency above approximately 1/3 of the maximum number of vCPUs. An economic analysis revealed that running on large scale cloud platforms is cost effective compared to traditional desktop systems.

**Conclusions**: Overall, cloud implementations of PhenoMeNal show excellent scalability for standard metabolomics computing tasks on a range of platforms, making them a compelling choice for research computing in metabolomics.

# Background

Metabolomics is the comprehensive study of metabolites, which are naturally occurring small molecules involved in almost all biological processes [1, 2]. It compliments other Big Data 'omics fields such as genomics and proteomics to facilitate a better understanding of biology at the molecular level across diverse areas of biology and medicine in areas such as disease pathology, drug discovery and toxicology. An increased awareness of the role of metabolomics has been fuelled by developments in medical science which has led some to consider it as a future driver of precision and personalized medicine [3].

In many respects, metabolomics is not as mature as other omics fields such as genomics or transcriptomics. Standards for sample collection, measurement and analysis have yet to be widely accepted and there is relatively limited use of commercial platforms. Given these experimental limitations, it is hard to develop computational approaches which are widely applicable and bespoke solutions are often required. There is therefore an acute need for standardised workflows (including scalable tools [4]) to increase throughput and reproducibility.

Over the last few years, the size of data sets generated from metabolomics studies has dramatically increased. An example of a source of metabolomic 'Big' data is the National Phenome Centre [5] where an average 1.5 Petabytes ($10^{15}$ B) of high quality metabolomic data can be produced every year from a wide variety of scientific and clinical studies. This dramatic increase in data generation has made it cumbersome to carry out extensive computational analyses using typical desktop machines. The trend towards larger, more heterogeneous datasets encourages a move towards scalable computation which can be provided by a cloud-based approach where compute and storage requirements match the computational power required.

Many of the analyses carried out in metabolomic studies are *embarrassingly parallel*; in other words, they can be easily broken down to small tasks that can be dealt with at the same time. To perform large processes concurrently in an efficient manner, multiple processors are required. These processors can be aggregated as individual central processing units (CPUs), cores or virtual machines. Given a parallel problem and a number of processors, *scalability* [6] is concerned with measuring the speedup gained as the number of processors increases. This focuses on the capacity of the parallel system to solve realistic problems, the ability to maintain a stable performance level and how effectively and cost-efficiently resources are utilised with increased capacity.

Cloud computing - the provision of infrastructure as a service - is one way to exploit parallel processing. Many areas in science, engineering and medicine are beginning to leverage cloud computing to store, process and deduce novel insights from large amounts of data [7]. Metabolomics can also benefit from cloud solutions, allowing users anywhere to access implementations that efficiently process large amounts of data, also facilitating easier collaboration between remote sites and introducing a standardised, reproducible and accepted data analysis process.

Here, we investigate how PhenoMeNal [8] (Phenome and Metabolome aNalysis), a European Union Horizon-2020 project that has developed a comprehensive and standardised e-infrastructure for analysing medical metabolic phenotype data, can perform in a variety of cloud infrastructures. The PhenoMeNal e-infrastructure is a scalable platform for the implementation of data processing and analysis pipelines of molecular phenotypic data. PhenoMeNal enables users to easily set-up and run metabolomics data processing on cloud resources. Users can deploy PhenoMeNal on several public and private clouds as well as on local servers, which improves maintainability as the software is designed to be cross-platform. The project has also embraced security and privacy (Ethical, Legal, and Social Implications (ELSI)). The tools provided by PhenoMeNal are packaged inside Docker containers [9]. This gives the environment the consistency and homogeneity required to run smoothly in different environments. To orchestrate these containers, PhenoMeNal employs Google's Kubernetes [10] as its cluster manager and the main tool used to deploy PhenoMeNal is KubeNow [11]. The portability of PhenoMeNal allows taking the compute to the data, a necessary strategy when dealing with confidential patient information in situations where the data is not fully anonymised. PhenoMeNal may be accessed here https://portal.phenomenal-h2020.eu/home .

PhenoMeNal provides many tools to process and analyse metabolomics data within highly parallelised cloud environments. Thus, important questions arise as to how typical PhenoMeNal tools scale with respect to the size of data sets and computational resources, and whether this provides real benefits (in terms of cost and efficiency) to users. We therefore set out to examine scalability of the PhenoMeNal infrastructure in this context. We deployed PhenoMeNal on four different computing platforms at different scales and assessed performance with two typical metabolomics tools – BATMAN for NMR data analysis and PAPY for power and sample size calculations.

## Data Description

For BATMAN, we used a large data set of 2000 1-dimensional CPMG $^1$H NMR spectra of blood serum from the MESA consortium [12, 13]. The spectral regions outside the region 0 to 10 ppm were removed from the data resulting in 68k data points per spectrum. The following BATMAN parameter values were used: 9 target metabolites, burn-in=1000, post burn-in=600, down sample size=10. In a typical BATMAN application, a range of parameter settings would be used, as iterative runs are used to optimize the fit. These settings correspond to those typically employed in the early stages of fitting.

For PAPY, we used 84 1-dimensional $^1$H NMR spectra of urine from the Airwave Study [14]. The spectral regions outside the region 0 to 10 ppm were removed and the data reduced by retaining one in every ten data points. The first 1000 variables were used. Parameter settings were as follows: both linear regression and classification models, 30 repeats at each grid point, sample size varied from 1 to 3000 in steps of 100, effect size varied from 0.05 to 0.75 in steps of 0.025.

The datasets used for testing BATMAN and PAPY are publicly available in https://doi.org/10.6084/m9.figshare.c.4204022 .

# Analyses

## Study Design

PhenoMeNal can be deployed on various cloud platforms as well as bare-metal environments. To examine scalability, we chose to investigate a range of compute scales (numbers of processors, memory available), including commercially available clouds as well as dedicated academic compute environments. The platforms used are a high performance Desktop PC, a local medium scale server, and the large scale Microsoft Azure and EMBASSY Clouds [15]. Figure 1 (in the Appendix) shows the test platforms and tools.

In computational studies, there are two scaling paradigms, namely algorithmic scaling and parallel scaling. In contrast to algorithm scaling, which focuses on measuring the increase in computational time as the size of computation increases, parallel scaling is concerned with measuring the decrease in wall clock time as more processors are used for performing calculations. In this work, we focus on parallel scaling experiments. There are two common parallel scaling experimental designs: strong scaling and weak scaling.

**Strong Scaling**: In this approach, the problem size (i.e. the size of the data) is fixed and the number of processors used to process the data is changed. Strong scaling is often used to examine problems that are CPU bound.

- **Weak Scaling**: Here, the problem size is changed as the number of processors is changed. The ratio of problem size to number of processors is kept constant. Weak scaling is often used to examine problems that are memory bound, since increasing the number of processors allows a problem to be addressed that could not fit in a single processor memory.

Scaling performance can be measured in terms of processing time, speed-up ratio and scaling efficiency (see Methods for details).

## Scaling of NMR Processing with the BATMAN Tool

Figure 2 (in the Appendix) illustrates the strong scaling performance with BATMAN. Performance improves dramatically moving from 1 vCPU (320,000 s) to 1000 vCPUs (563 s), a speed up of 568 times. The practical advantage of such a change is clear, converting a run of approximately 4 days to around 10 min. It is interesting to note that all systems show very similar performance when evaluated at the same scale, indicating relatively little influence from the large differences in architecture. Each platform exhibits a fairly consistent linear speedup across 3 orders of magnitude in compute resources. This is perhaps surprising, considering the overhead resulting from network traffic, load balancing, disk IO and so on. However, it is noteworthy that the speedup tends to plateau at ~800 to 1000 vCPUs on EMBASSY and Azure respectively. All tested cloud platforms maintain good efficiency of ≥80% up to approximately one third of the maximum available vCPU resources (~300 vCPUs on EMBASSY and Azure).

The results of weak scaling tests with BATMAN are illustrated in Figure 3 (in the Appendix). Speed up is not shown as it is not defined for weak scaling. The runtime on all four systems in the weak scaling regime is surprisingly similar at low and mid ranges. Runtimes remain fairly flat until a limit at which they rapidly increase. This increase begins at approximately one third of the available vCPUs on Azure and EMBASSY, but at a higher proportion of available resources on the Desktop and Local Server. The limit can perhaps be attributed to the large number of workers generating significant network overhead and job orchestration requirements. Increases in runtime lead to a drop in weak scaling efficiency. This is similar to the efficiency drop observed in the strong scaling tests, suggesting that for BATMAN, the problem is not purely memory limited.

## Scaling of the Power Analysis Tool – PAPY

Figure 4 (in the Appendix) and Figure 5 (in the Appendix) show the results for strong and weak scaling with the power analysis tool. Due to extremely lengthy runtimes, runs with one vCPU, $n=1$, were not performed. Instead, $T_1$ was estimated using the lowest number of vCPUs tested on each platform using the following equation:

$$T_1 = n'T_{n'}$$

where $n'$ is the lowest number of vCPUs used and $T_{n'}$ is the time taken for this number of vCPUs. This estimated $T_1$ was then used to calculate the speedup and efficiency using equations (1-3). Hence, the points starting each series in Figure 4a (in the Appendix) lie on the diagonal (linear scaling) line.

In the strong scaling regime, Figure 4 (in the Appendix) illustrates that PAPY scaled almost linearly across the three orders of magnitude of computational resources. For all systems, a minor levelling off was observed around a third to a half of the maximum resources (i.e.300-500 vCPUs on Azure and EMBASSY). This translated to a decline in efficiency to around 80%. The 8-core desktop PC took about 25 days to run PAPY. This was reduced to around 30 minutes on EMBASSY at 1000 vCPUs, clearly an extremely advantageous time saving which would enable extensive investigation of the parameter space for this dataset.

Weak scaling with PAPY showed a slightly poorer performance than that observed with BATMAN, as shown in Figure 5 (in the Appendix). On the desktop, the run time increased from approximately 120 s at 2 CPUs to around 370 s at 8 CPUs reflecting a dramatic drop in efficiency to near 30%. However, on the larger systems run time started low (about 30 s at 10 vCPUs) and increased only gradually, with a slight upturn at the largest job sizes. The scaling efficiency dropped steadily, reaching 50% at about 500 vCPUs for Azure and EMBASSY. The local server appeared to show steeper scaling behaviour, dropping to near 65% at 50 vCPUs. All systems showed steeper efficiency curves than with BATMAN. This may be due to differences between the two tools in the ratio of compute to overhead when creating, loading, monitoring and finishing a large number of jobs. This might be improved for PAPY by increasing the computational intensity of each job, i.e. increasing the size of the sample, the size of effect, and the number of repeats.

## Discussion and Implications

To our knowledge, this is the first assessment of the computational scalability of metabolomics software tools on large scale compute platforms. By running the tools across four platforms of increasing computational scale, we have been able to assess scalability not only across 3 orders of magnitude, but also in a variety of diverse computational environments. This increases the reliability, robustness and reproducibility of these results. A tutorial on the scalability experiments is available at:

https://github.com/csmsoftware/phnmnl-scalability

The main finding is that large jobs using established metabolomics tools can be run in a compute and time efficient manner in the PhenoMeNal environment. This is despite the fact that the setup of the environment, although straightforward for a skilled computational scientist, may not be as straightforward for a domain specialist. To assist users, PhenoMeNal may be deployed on several cloud platforms with comprehensive guidance available from the PhenoMeNal website (https://portal.phenomenal-h2020.eu/home).

Surprisingly, the scaling behaviour of both tools examined was unexpectedly linear, especially in the strong scaling regime. As expected, loss of efficiency was found for large numbers of vCPUs, although this was significantly better than reported in previously published examples from other domains [16]. In addition, weak scaling efficiency was somewhat poorer for the power analysis tool PAPY than for BATMAN. It is possible that the relative computational complexity of the two algorithms influences this behaviour as PAPY scaling was improved when using parameter settings leading to computationally intensive runs. It was also apparent that similar scaling performance can be achieved on both a local medium scale server and a cloud-based system.

In these experiments we found that Luigi can be a limiting factor in scaling large jobs. As the number of Luigi workers increases, the system can become unresponsive and tasks can be frozen or terminated by the operating system. We believe this is because Luigi is designed for managing batch workflows, rather than as a platform for running large numbers of tasks in parallel. If more resources are allocated to the machine running Luigi, performance improves. For example, when running Luigi using 12 vCPUs, the highest number of parallel jobs it was able to manage was 200; when we increased the number of vCPUs to 20, it was able to run up to 300 parallel jobs. It is possible that platforms demonstrating better scaling characteristics for parallel jobs, such as Apache Spark [17] or NextFlow [18] may be more appropriate for large scale parallel job control.

While the speed improvements of running on a large number of processors are clear, one may ask whether these are reflected in high running costs as compared to a typical desktop system. To process 2000 spectra processed with BATMAN on the Azure Platform with 300 vCPUs, which at the time of writing cost ca. $16.59 per hour, the running costs will be less than $10, including system deployment (about 10 min, ~$2.77) and runtime (about 22 min, ~$6.16 ). This can be compared with the required runtime (about 16 hours) on a Desktop PC (Intel i7, 8 cores 3.4GHz, 16GB RAM) with a capital cost of ~$1500 and further running costs (maintenance, energy, and support *etc*.). Therefore, when considering equivalent large processing jobs, the cloud-based systems are, at the least, competitive with traditional desktop systems. They have the economic advantages of being affordable, as well as avoiding maintenance, support and space costs and can enable careful exploration of the parameter space thus potentially improving the quality and testing of models.

## Methods

### Platforms and Tools

The detailed specifications of each of the platform are as follows:

- **Desktop PC** - Intel i7 CPU with 3.4GHz and 8 cores, 16GB RAM. Run the code directly.

- **Local Server** - This server is equipped with 40 CPUs (with 2 cores per CPU providing 80 cores), and 1TB RAM. The bare-metal version of PhenoMeNal was deployed creating a virtual cluster of one large VM with 70 virtual CPUs (vCPUs) and 300GB of RAM. Due to job design, a maximum of 50 vCPUs were available for compute jobs, as detailed in Table 1. This server is part of the MedBio architecture [19] at Imperial College London.

- **Microsoft Azure** - This system provided 1200 virtual CPUs (a maximum of 1000 vCPUs were available for compute jobs) spread over 36 VMs, 4.5TB RAM in total. PhenoMeNal was deployed on a cluster with specifications detailed in Table 1.

- **EMBASSY Cloud** - hosted at EMBL-EBI in Hinxton, Cambridge, UK. EMBASSY cloud is based on OpenStack [20], which is an open source cloud operating system. The tenancy provides 1440 vCPUs (with a maximum of 1000 vCPUs for compute jobs) and 2.9TB RAM that can spread over 57 VMs. A cluster was created with the specifications detailed in Table 1.

*Table 1:* KubeNow set up on Azure and EMBASSY Clouds

| Node type | Count | Total No of vCPUs | RAM in GB |
|---|---|---|---|
| **Local Server*** | | | |
| Master | 1 | 2 | 4 |
| Worker Nodes | 7 | 50 | 300 |
| **Azure** | | | |
| Master | 1 | 2 | 7 |
| GlusterFS | 1 | 8 | 32 |
| Worker Nodes | 34 | 1000 | 4352 |
| **EMBASSY** | | | |
| Master | 1 | 22 | 36 |
| GlusterFS | 1 | 4 | 8 |
| Worker Nodes | 55 | 1000 | 1980 |

*Bare-metal deployment on local server does not require specification of the number of GlusterFS nodes.

The tests were conducted using two PhenoMeNal tools with two types of tests for each tool. The two tools were BATMAN [21], and PAPY [22].

BATMAN is an R-based package for deconvolving and quantifying peaks from small molecules in 1D NMR spectra of complex mixtures. It employs a spectral library to fit the characteristic peak patterns of known compounds and is able to deal with shifts in peak positions between spectra and highly overlapped resonances. The model is solved using a Bayesian Markov Chain Monte Carlo approach which is computationally intensive, typically requiring several hours for a single test spectrum on a high-performance desktop machine. The main output is a set of estimated relative concentrations for each target metabolite in each test spectrum.

PAPY is a Python based package for estimating statistical power and sample size in metabolomics studies. It employs a pilot data set to simulate data with varying sample and effect sizes. The model computes log-normal distributions and incorporates correlation between variables. The compute time depends primarily on the number of variables in the pilot study and the granularity of the sample size - effect size grid, and can take several hours on a standard desktop machine.

Parallel Scaling

Two common methods of summarising scaling performance are *Speedup* and *Scaling Efficiency* [23]. Let us assume that we have a strong scaling case. We denote the time taken to solve a problem using the most basic unit of processing (e.g. one processor) by $T_1$ and the time taken to solve the same problem using *n* identical processors by $T_n$. Then Speedup *S(n)*, gained by using *n* processors, is calculated as:

$$S(n) = \frac{T_1}{T_n} \qquad (1)$$

Therefore, we can define Speedup as the ratio of the time taken to solve a problem using a single processor to the time taken to solve the same problem by several identical processors that are used at the same time (*i.e. in parallel*). For strong scaling, the efficiency can be calculated as:

$$E(S) = \frac{T_1}{nT_n} = \frac{S(n)}{n} \qquad (2)$$

Hence, the efficiency can be defined as the ratio of Speedup when solving a problem using *n* processors to the number of processors itself. If increasing the number of processors by a factor *k* leads to a speedup by the same factor *k* then the system is said to scale linearly and the strong scaling efficiency is 100%. If, however, the speedup is less than the increase in the number of processors, then the scaling efficiency is less than 100%.

In the weak scaling case, the problem size grows with the number of processors. Hence, if the time taken for one processor is *T'₁* and that for *n* processors is *T'ₙ*, then the weak scaling efficiency can be calculated as:

$$E(W) = \frac{T'_1}{T'_n} \qquad (3)$$

Thus, if increasing both the number of processors and job size by a factor *k* results in the same computation time, the weak scaling efficiency is 100%. If the computation takes longer, the efficiency falls below 100%. Note that speed up is not defined for weak scaling.

In our tests, speed up and scaling efficiency were calculated for each system separately, by normalising to the relevant $T_n$ of each system.

## Parallel Job Control

PhenoMeNal tools can be accessed via three interfaces: Galaxy, Jupyter notebooks, and Luigi [24]. Our tests used Luigi, a Python package for constructing workflows that consist of one or more tasks, controlled and monitored using a web interface. When a workflow is run, it is handled by a Luigi worker. PhenoMeNal developed a module that enables running Luigi workers on Kubernetes [25], implemented here by running a worker inside a Kubernetes pod. The module enables setting up the resources that can be used by each Luigi worker, running several identical workers in parallel.

## Experimental Setup

As assessment of strong and weak scaling was the main focus of this work, the aim was to ascertain how processing times change with the size of data set, the number of parallel tasks (or jobs), the number of processors and additional parameters. BATMAN is designed to run on a multi-processor machine. In order to run it in parallel on a computer cluster, we took advantage of the fact that BATMAN can handle separate spectra independently. Partitioning datasets to make the best use of cloud servers is an NP-hard problem [26] and not the focus of this paper. We therefore choose the following simple scheme to partition these datasets. The 2000 spectra were split into 200 files (10 spectra per file). The objective is to have multiple instances of BATMAN running at the same time (on the same or different nodes in the cluster). A complete guide on how to run similar experiments on the PhenoMeNal infrastructure can be found in [27]. For PAPY, a similar approach was taken. The pilot data was split into 10 variables per file, resulting in 100 files in total. Hence, each Luigi worker processed 10 variables using 10 vCPUs in parallel. Similar to BATMAN, the PAPY tool was dockerised and run as Kubernetes pods via Luigi workers.

Python scripts were used within PhenoMeNal's Jupyter interface to configure Luigi, following the instructions provided in [28]. This enabled control of parallel workers on the cluster, specification of resources available to each worker, inspection of worker status etc. BATMAN and PAPY were tested separately. When submitting jobs to the cluster, each job was hosted by a pod, and each pod run as a Luigi worker. Hence, one node in the cluster could run one or more Luigi workers (and consequently one or more BATMAN/PAPY job instances). The resources allocated to each Luigi worker were 10 vCPUs and 16GB RAM. Note that we did not use exactly 100% of the maximum number of vCPUs on the local server, Azure or EMBASSY, since some vCPUs were required for other purposes (e.g. job control, file server *etc*.). All tests were repeated 3 times.

The number of vCPUs was varied by changing the number of Luigi workers. In summary:

- **Strong scaling**: The size of the problem was fixed (BATMAN - 2000 spectra, PAPY - 1000 variables) and the number of vCPUs varied from 1 to 1000 across the 4 systems.

- **Weak scaling**: The size of problem and number of vCPUs were both varied, maintaining a constant ratio of the size of the problem to the number of vCPUs. For BATMAN, we used 1 vCPU per NMR spectrum. For PAPY, we used 1 vCPU per variable.

# Appendix

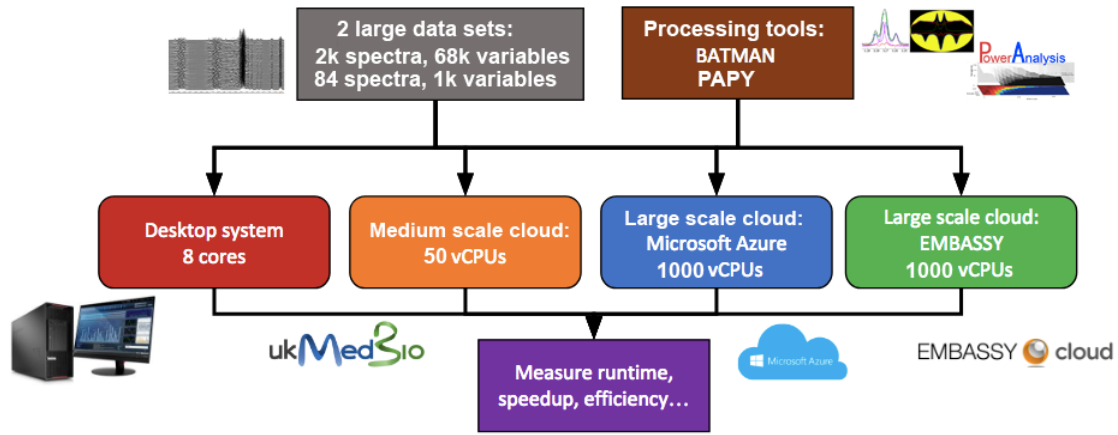

Figure 1: Scalability set up. Two datasets and two metabolomics tools are run on four compute architectures. In each case performance is measured in terms of run time, speed up and scaling efficiency.

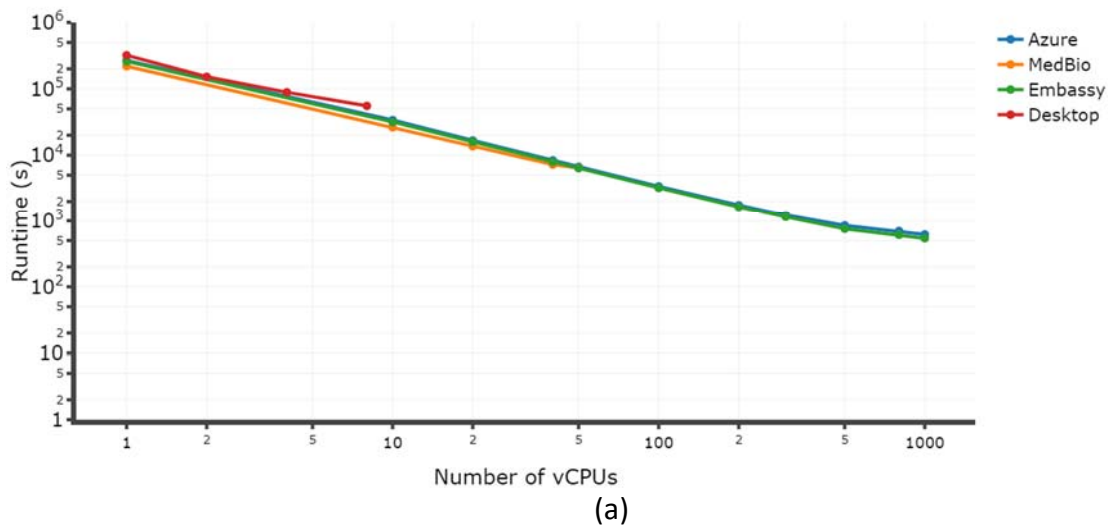

(a)

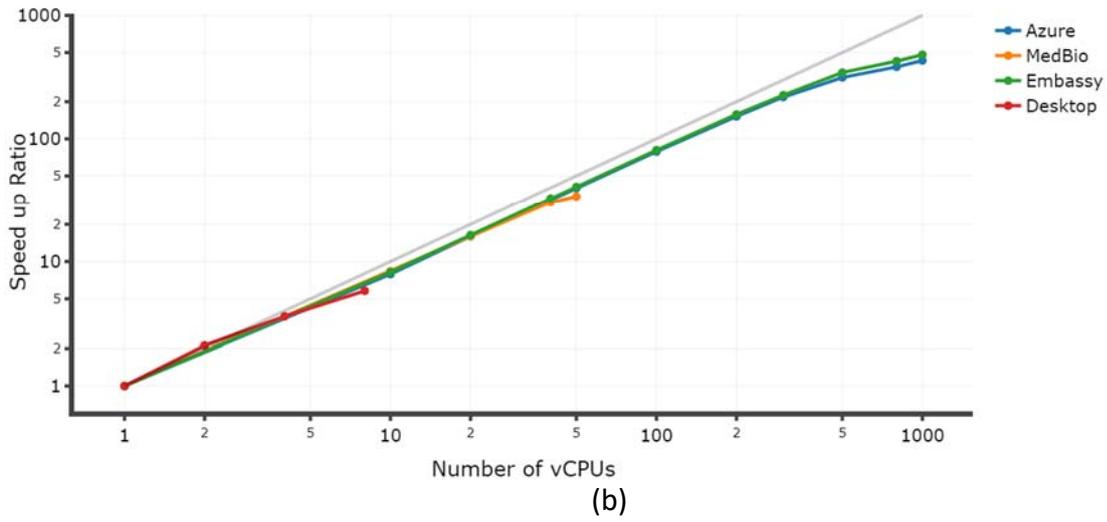

(b)

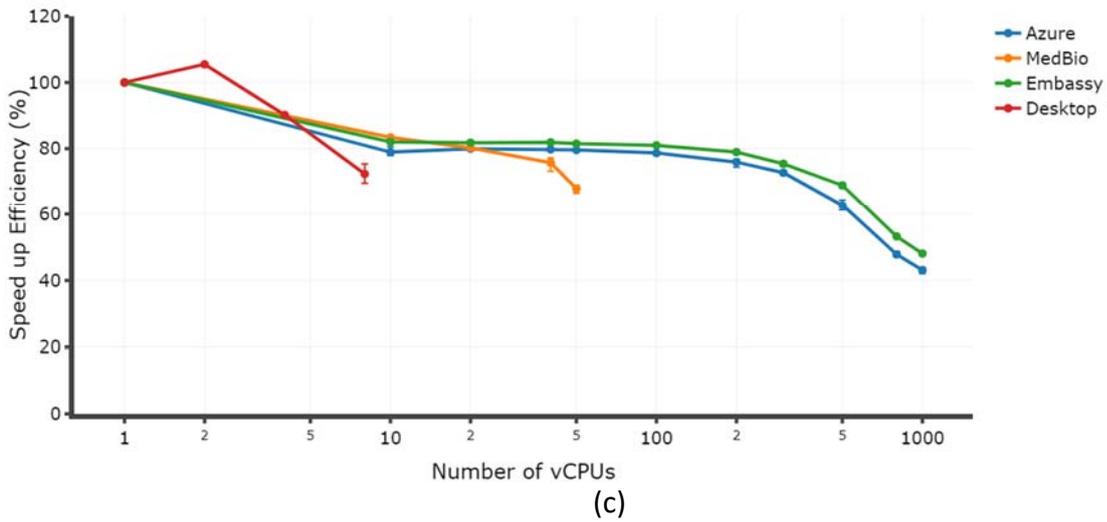

(c)

Figure 2: BATMAN Strong Scaling Performance. (a) Runtime (b) Speedup ratio (normalized to $T_1$ of each platform). Grey diagonal line represents the ideal (linear) speedup ratio. (c) Strong Scaling Efficiency (normalized to $T_1$ of each platform). Note log scales. Points show mean values across 3 replicate experiments; error bars show minimum and maximum. In some cases, error bars are too small to be visible.

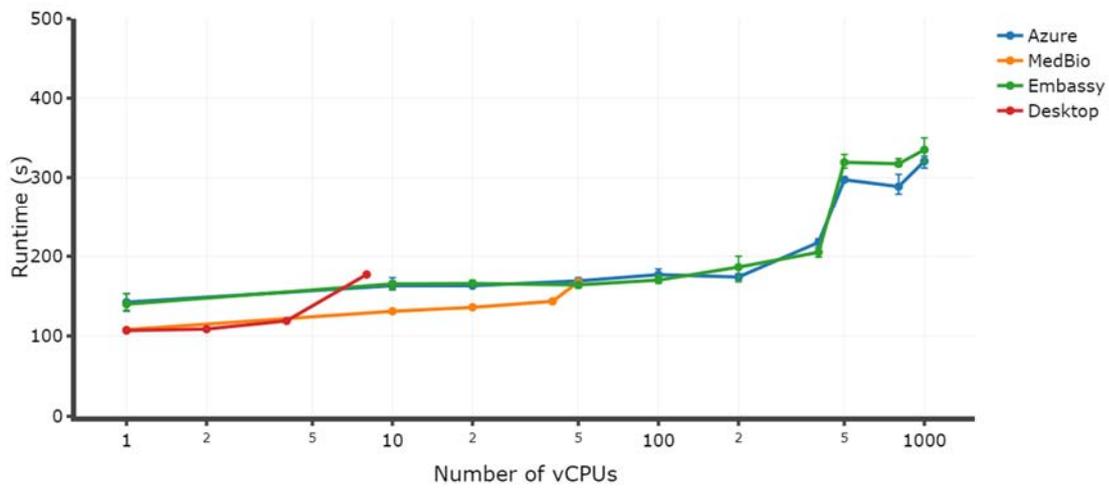

(a)

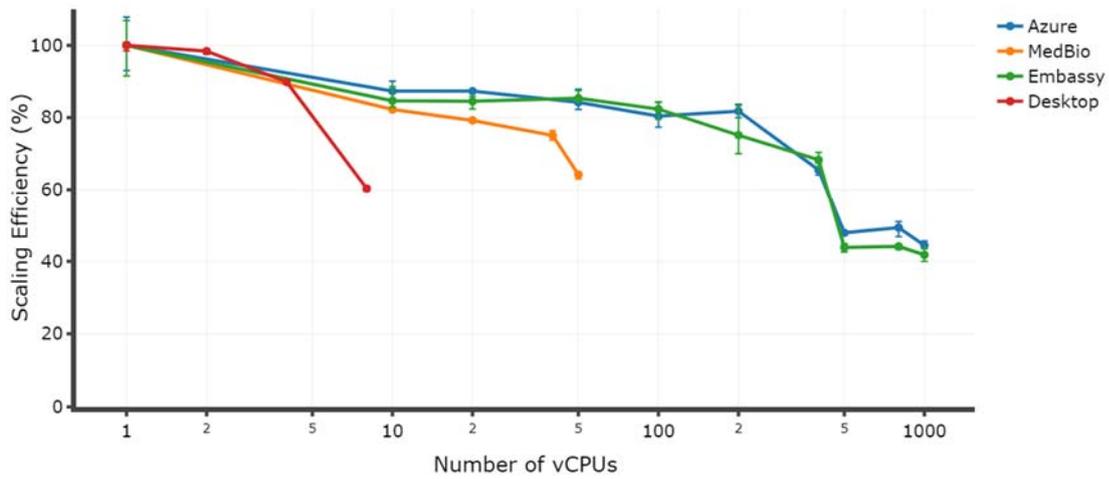

(b)

Figure 3: BATMAN Weak Scaling Performance. (a) Runtime (b) Weak Scaling Efficiency. Points and error bars as for Figure 2.

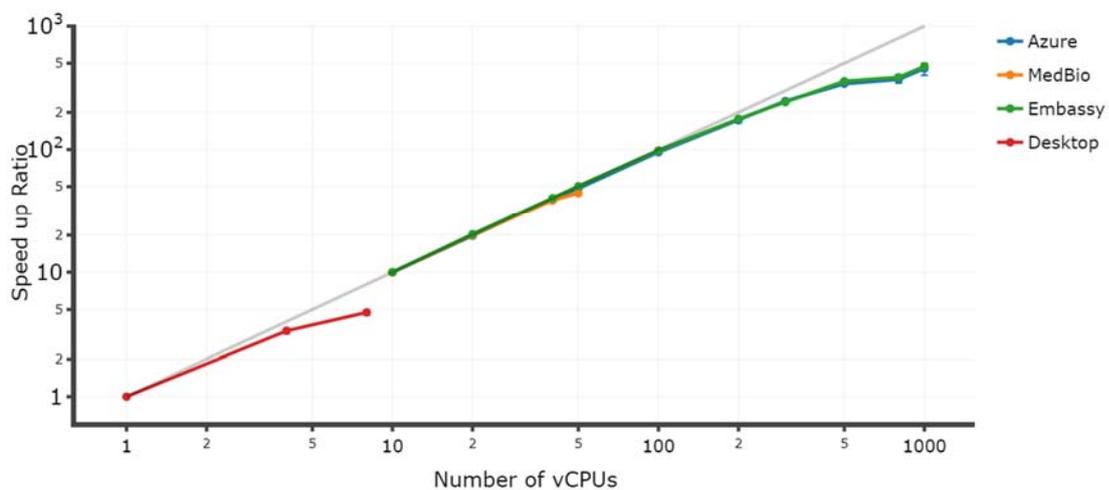

(a)

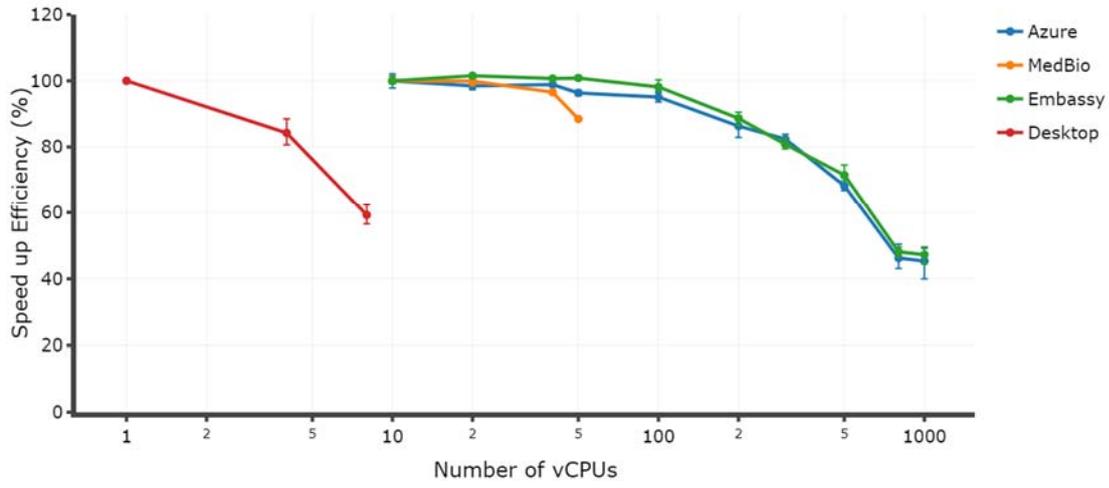

(b)

Figure 4: PAPY strong scaling performance. (a) Strong scaling speedup ratio (normalised to lowest $nT_n$ of each platform). (b) Strong scaling efficiency. Points and error bars as for Figure 2.

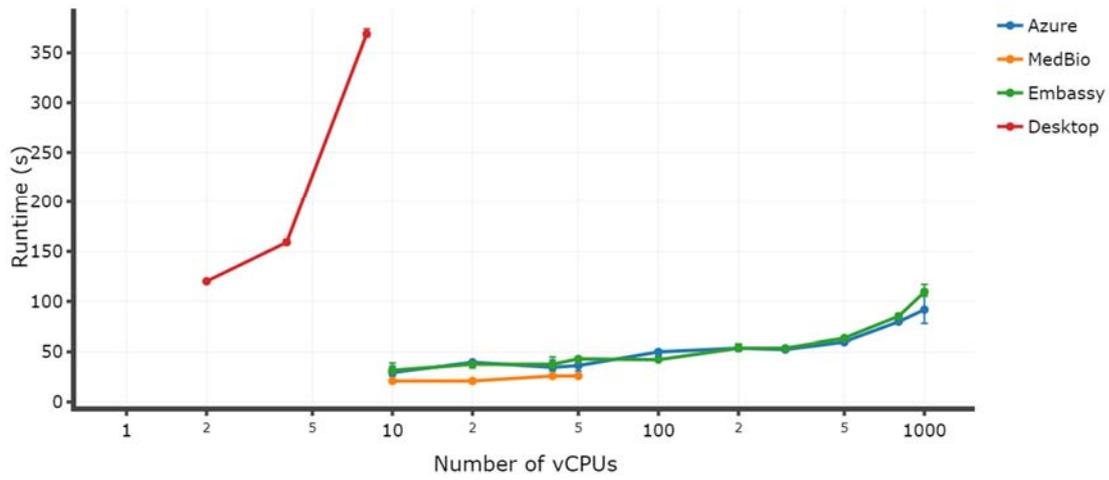

(a)

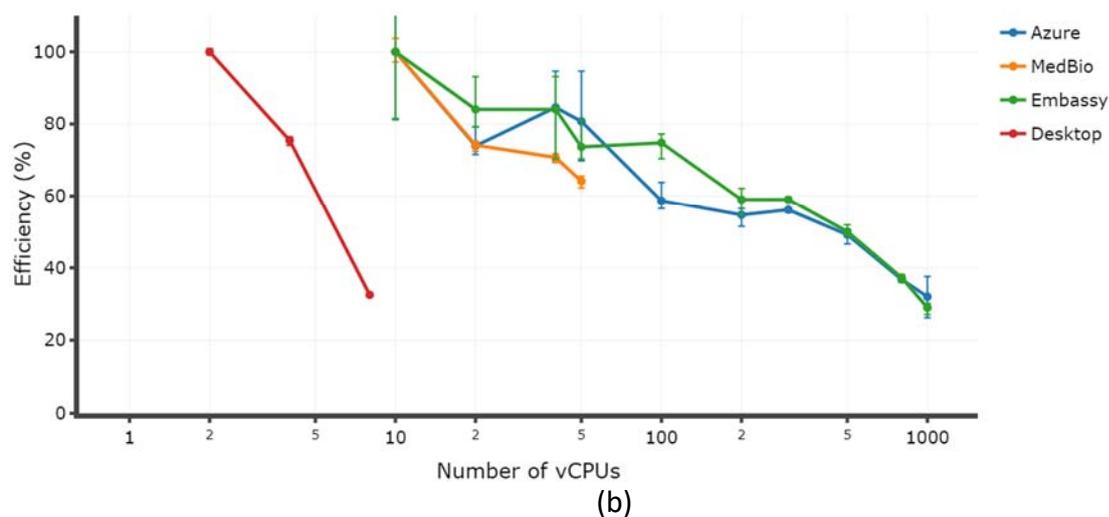
(b)

Figure 5: PAPY weak scaling performance. (a) Weak scaling runtime. (b) Weak scaling efficiency. Points and error bars as for Figure 2.

# Acknowledgement
This project was funded by the European Commission PhenoMeNal Grant EC654241.

**REFERENCES**

1.      Nicholson JK, Lindon JC, Holmes E. 'Metabonomics': understanding the metabolic responses of living systems to pathophysiological stimuli via multivariate statistical analysis of biological NMR spectroscopic data. Xenobiotica. 1999;29(11):1181.

2.      Fiehn O. Metabolomics--the link between genotypes and phenotypes. Plant Mol Biol. 2002;48(1-2):155-71.

3.      Trivedi DK, Hollywood KA, Goodacre R. Metabolomics for the masses: The future of metabolomics in a personalized world. New Horiz Transl Med. 2017;3(6):294-305.

4.      Sarkar BK. Big data for secure healthcare system: a conceptual design. Complex & Intelligent Systems. 2017;3(2):133-51.

5. Saini A. London's Olympic Drug Testing Lab to Become National Phenome Center. Science. 2012;337(6094):513.

6. Kumar VP, Gupta A. Analyzing Scalability of Parallel Algorithms and Architectures. Journal of Parallel and Distributed Computing. 1994;22(3):379-91.

7. Hashem IAT, Yaqoob I, Anuar NB, Mokhtar S, Gani A, Khan SU. The rise of "big data" on cloud computing: Review and open research issues. Inform Syst. 2015;47:98-115.

8. Peters K, Bradbury J, Bergmann S, Capuccini M, Cascante M, de Atauri P, et al. PhenoMeNal: Processing and analysis of Metabolomics data in the Cloud. GigaScience. 2018:giy149-giy.

9. Merkel D. Docker: lightweight Linux containers for consistent development and deployment. Linux J. 2014;2014(239):2.

10. Hightower K, Burns B, Beda J. Kubernetes: Up and Running: Dive Into the Future of Infrastructure: " O'Reilly Media, Inc."; 2017.

11. Capuccini M, Larsson A, Carone M, Novella JA, Sadawi N, Gao J, et al. KubeNow: an On-Demand Cloud-Agnostic Platform for Microservices-Based Research Environments. arXiv preprint arXiv:180506180. 2018.

12. Bild DE, Bluemke DA, Burke GL, Detrano R, Diez Roux AV, Folsom AR, et al. Multi-Ethnic Study of Atherosclerosis: objectives and design. Am J Epidemiol. 2002;156(9):871-81.

13. Karaman I, Ferreira DL, Boulange CL, Kaluarachchi MR, Herrington D, Dona AC, et al. Workflow for Integrated Processing of Multicohort Untargeted 1H NMR Metabolomics Data in Large-Scale Metabolic Epidemiology. J Proteome Res. 2016;15(12):4188-94.

14. Elliott P, Vergnaud A-C, Singh D, Neasham D, Spear J, Heard A. The Airwave Health Monitoring Study of police officers and staff in Great Britain: Rationale, design and methods. Environ Res. 2014;134:280-5.

15. Cook CE, Bergman MT, Finn RD, Cochrane G, Birney E, Apweiler R. The European Bioinformatics Institute in 2016: Data growth and integration. Nucleic Acids Res. 2016;44(D1):D20-D6.

16. Perlin N, Zysman JP, Kirtman BP. Practical scalability assesment for parallel scientific numerical applications. arXiv preprint arXiv:161101598. 2016.

17. Zaharia M, Xin RS, Wendell P, Das T, Armbrust M, Dave A, et al. Apache spark: a unified engine for big data processing. Communications of the ACM. 2016;59(11):56-65.

18. Di Tommaso P, Chatzou M, Floden EW, Barja PP, Palumbo E, Notredame C. Nextflow enables reproducible computational workflows. Nat Biotechnol. 2017;35(4):316.

19. MedBio - UK Medical Bioinformatics Partnership Programme. http://www.imperial.ac.uk/uk-med-bio. Accessed 05 Jun 2018.

20. Sefraoui O, Aissaoui M, Eleuldj M. OpenStack: toward an open-source solution for cloud computing. International Journal of Computer Applications. 2012;55(3):38-42.

21. Hao J, Astle W, De Iorio M, Ebbels TM. BATMAN--an R package for the automated quantification of metabolites from nuclear magnetic resonance spectra using a Bayesian model. Bioinformatics. 2012;28(15):2088-90.

22. Blaise BJ, Correia G, Tin A, Young JH, Vergnaud AC, Lewis M, et al. Power Analysis and Sample Size Determination in Metabolic Phenotyping. Anal Chem. 2016;88(10):5179-88.

23. Sahni O, Carothers CD, Shephard MS, Jansen KE. Strong scaling analysis of a parallel, unstructured, implicit solver and the influence of the operating system interference. Scientific Programming. 2009;17(3):261-74.

24. Luigi. https://github.com/spotify/luigi. Accessed 08 Feb 2019.

25. Saito H, Lee H-CC, Hsu K-JC. Kubernetes Cookbook: Packt Publishing Ltd; 2016.

26. Vaquero LM, Celorio A, Cuadrado F, Cuevas R. Deploying large-scale datasets on-demand in the cloud: treats and tricks on data distribution. IEEE Transactions on Cloud Computing. 2015;3(2):132-44.

27. A Tutorial on the Scalability of the PhenoMeNal Project. https://github.com/csmsoftware/phnmnl-scalability. Accessed 08 Feb 2019.

28. An OpenMS preprocessing workflow and R downstream analysis using PhenoMeNal's Jupyter. https://github.com/phnmnl/MTBLS233-Jupyter. Accessed 08 Feb 2019.